\documentclass[
    aps,prb,twocolumn,groupedaddress,
    superscriptaddress,amsfonts,
    amssymb,amsmath,citeautoscript,
    longbibliography,a4paper
    ]{revtex4-2}

\usepackage[utf8]{inputenc}
\usepackage[english]{babel}

\usepackage{microtype}
\usepackage{xspace}

\usepackage{txfonts}  
\usepackage{txfontsb}

\usepackage{bm} 

\usepackage{xcolor}
\usepackage[]{graphicx}

\usepackage[]{booktabs}
\usepackage{array}
\usepackage{layouts}
\usepackage{multirow}

\usepackage{enumerate}
\usepackage[inline]{enumitem}

\usepackage{hyperref}
\hypersetup{colorlinks,
	linkcolor={blue!75!black!80!yellow},
	citecolor={blue!75!black!80!yellow},
	urlcolor={blue!75!black!80!yellow}
}

\usepackage[capitalize,nameinlink]{cleveref}

\crefname{subequations}{Eqs.}{Eqs.} %
\Crefname{subequations}{Eqs.}{Eqs.}
\crefformat{subequations}{#2Eqs.~(#1)#3}
\Crefformat{subequations}{#2Eqs.~(#1)#3}
\crefname{page}{p.}{p.} 

\usepackage{placeins}

\usepackage{siunitx}
\DeclareSIUnit[number-unit-product = ]\percent{\char`\%} 

\usepackage[centering,hmargin=16mm,tmargin=30mm,bmargin=26mm]{geometry}

\renewcommand{\paragraph}[1]{\vskip 1ex\noindent\textbf{#1.}~}

\usepackage[eulergreek]{sansmath}
\makeatletter
\renewcommand\@make@capt@title[2]{%
    \@ifx@empty\float@link{\@firstofone}{\expandafter\href\expandafter{\float@link}}%
    \sansmath\sffamily\textbf{#1\@caption@fignum@sep}#2 
}%
\renewcommand\figurename{Figure}
\makeatother
\usepackage{svg}
\usepackage{placeins}
\newcommand{\rv}{\mathbf{r}}

\newcommand{\Ev}{\mathbf{E}}

\newcommand{\Ef}{E_{\text{F}}}
\newcommand{\kb}{k_{\text{B}}}

\newcommand{\dd}{\mathrm{d}}

\newcommand{\dde}{\,\dd}
\newcommand{\iu}{\mathrm{i}}
\newcommand{\ie}{i.e.,\@\xspace} 
\newcommand{\cf}{cf.\@\xspace}

\newcommand{\appropto}{\mathrel{\vcenter{
			\offinterlineskip\halign{\hfil$##$\cr
				\propto\cr\noalign{\kern.2pt}\sim\cr\noalign{\kern-2.5pt}}}}}

\let\Re\relax 
\DeclareMathOperator{\Re}{Re}
\let\Im\relax 
\DeclareMathOperator{\Im}{Im}

\graphicspath{{figs/}}
\begin{document}
\title{Adaptive multi-spectral mimicking with 2D-material nanoresonator networks}


\newcommand{\umnaffil}{Department of Mechanical Engineering, University of Minnesota, Minneapolis, MN 55455, USA}
\newcommand{\dtuaffil}{Department of Electrical and Photonics Engineering, Technical University of Denmark, 2800 Kgs.\ Lyngby, Denmark}

\author{Yujie Luo}
\affiliation{\umnaffil}

\author{Thomas Christensen}
\affiliation{\dtuaffil}

\author{Ognjen Ilic}
\affiliation{\umnaffil}


\begin{abstract}
    Active nanophotonic materials that can emulate and adapt between many different spectral profiles---with high fidelity and over a broad bandwidth---could have a far-reaching impact, but are challenging to design due to a high-dimensional and complex design space. Here, we show that a metamaterial network of coupled 2D-material nanoresonators in graphene can adaptively match multiple complex absorption spectra via a set of input voltages. To design such networks, we develop a semi-analytical auto-differentiable dipole-coupled model that allows scalable optimization of high-dimensional networks with many elements and voltage signals. As a demonstration of multi-spectral capability, we design a single network capable of mimicking four spectral targets resembling select gases (nitric oxide, nitrogen dioxide, methane, nitrous oxide) with very high fidelity (${>}\,90\%$). Our results are relevant for the design of highly reconfigurable optical materials and platforms for applications in sensing, communication and display technology, and signature and thermal management.
\end{abstract}

\maketitle


\section{Introduction}

The principles of metamaterial optics have made it possible to realize complex spectral responses that are not found in conventional materials. The burgeoning field of optical metamaterials, including artificial structures like dielectric metalenses, photonic crystal thermal emitters, or signal-processing metasurfaces, has inspired numerous applications across optical and photonic technologies~\cite{Chen2016-va,Urbas2016-lu,Khorasaninejad2017-hc,Quevedo-Teruel2019-ye,Ren2020-bj,He2022-ue}.  For the most part, such structures are static. They are often designed to have a target spectral response and, once fabricated, their properties cannot be changed. Going beyond such single-spectrum metamaterials, a broader set of questions emerges: to what extent can a single device or structure mimic and reconfigure between a range of complex spectral targets? What are the optical design principles to make such structures, and, are these principles specific to given targets or can they be made universal?

Active and reconfigurable optical platforms---structures and devices whose optical behavior can be dynamically tuned---could play a pivotal role in a broad range of applications~\cite{Shaltout2019-ip,Abdollahramezani2020-xv,Shalaginov2020-yn,Badloe2021-zq,Malek2021-mi,Morsy2021-pv,Zhu2021-tg,Tong2015-jj}, but designing a single structure that can accurately emulate many different spectra poses a complex design challenge. The root of the complexity lies in a parameter space that is inherently high dimensional. For a reconfigurable metamaterial or metasurface, the combination of a continuum of allowable geometrical parameters (i.e., nano-structure shape, size, and dimensionality) and material properties (i.e., refractive index) make the design space large and extracting intuition difficult. The unit cell can assume a range of topologies and symmetries in one, two, or three dimensions, and optical properties can be tuned in various ways, such as by mechanical deformation~\cite{Liu2017-yx,Arbabi2018-fz,She2018-xu}, electrostatic and electrochemical effects~\cite{Sherrott2017-rt,Kafaie_Shirmanesh2018-ij,Ding2018-uu,Di_Martino2016-ph,Zanotto2017-oy}, and material phase changes~\cite{Wang2015-hw,Rensberg2016-mz,Coppens2017-np,Zhang2021-pk,Fang2021-dh,Wu2017-ip,Zhang2021-ee}. Presently, designing active devices with multi-spectral capabilities in such a complex parameter space is often conceptually challenging and computationally demanding.

\begin{figure}[b]
    \centering%
    \includegraphics[scale=1]{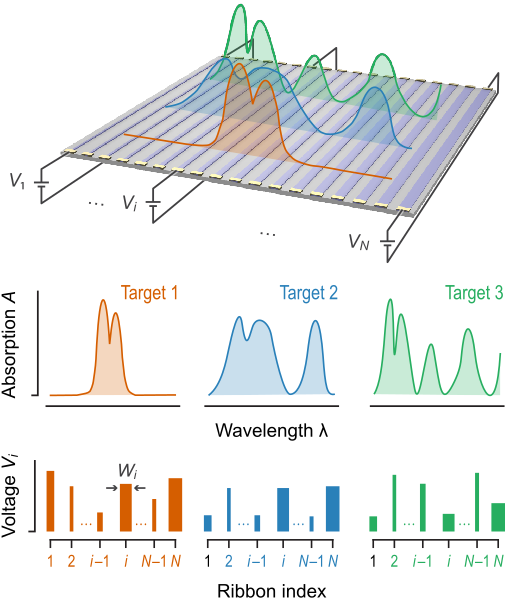}
    \caption{%
        \textbf{Multi-spectral photonic reconfigurability.}
        A network of coupled elements in a single device adaptively mimics multiple complex spectral profiles under a controlled external input.
        For a network of nano-structured graphene ribbons of optimized sizes $W_i$, applied voltage profiles $V_i$ switch between multiple absorptivity spectra (targets 1--3).
    }
    \label{fig:1}
\end{figure}

Here, we analyze the multi-spectral capability of a 2D-material nanoresonator network in graphene to adapt between a range of spectral targets. As shown in \cref{fig:1}, the network consists of optically coupled nanoribbons whose response depends on parameters set during fabrication (such as element size and arrangement in the network) and parameters that can be tuned post-fabrication (such as voltage).
Graphene, being a 2D semi-metal, has optical properties that can be strongly influenced by electric fields, leading to its use in tunable applications such as electronic modulation of emission, infrared absorption, and amplitude and phase modulation~\cite{Wang2012-wf,Yao2014-cq,Brar2015-fo,Li2016-cd,Ilic2018-ue,Han2020-pe,Khaliji-Low:PhysRev20-i, Nagpal2022-ew}. 

The goal of this work is to explore the potential of a single graphene network to reconfigure between multiple complex spectral targets, such as those found in the spectral absorption features of select gases. Absorption spectra of gases represent useful spectral targets since they are characterized by both narrow resonances and broader spectral envelopes. We approach the problem of reconfigurable spectral mimicking in several steps: first, we develop and validate a semi-analytical dipole-coupled model that is numerically efficient; second, we implement the model to make it compatible with modern auto-differentiation techniques, ensuring efficient optimization of high-dimensional networks with many elements; next, we use this approach to design a network that can mimic a spectrum resembling nitric oxide (NO) with high fidelity, in both a planar and a stratified configuration; finally, we demonstrate multi-spectral reconfigurability by designing a single network that can reconfigure between complex spectra resembling four gases. 

\section{Methods}
Nanostructured plasmonic 2D elements, such as individual graphene ribbons, exhibit a resonant modal response that is governed by the element's size (ribbon width, $W$) and material response (conductivity, $\sigma$) that is captured by a 2D polarizability~\cite{abajo2015plasmonics,christensen2017from}:
\begin{equation}
    \alpha(\omega) = 2W^3 \sum_n \frac{A_n}{\zeta_n - \zeta(\omega)},
\end{equation}
where $A_n$ are dimensionless mode- and polarization-dependent oscillator amplitudes, $\zeta_n$ are dimensionless mode-dependent eigenvalues, and $\zeta(\omega) = 2\iu\varepsilon_0\varepsilon_{\text{B}}\omega W/\sigma(\omega)$ captures the frequency ($\omega$), size ($W$), and response ($\sigma$) dependence (with vacuum $\varepsilon_0$ and relative background permittivities $\varepsilon_{\text{B}}$).
The coefficients $A_n$ and $\zeta_n$ depend solely on shape---but not the size or material---of the nanostructured element.
For a ribbon, the quantum numbers $n$ implicitly include the momentum along the ribbon: for normal incidence, we consider the zero-momentum limit where the dipole resonance is characterized by $\zeta_1 \approx 2.3159$ and $A_1 \approx 0.8791$.
In total, we include the first ``bright'' 6 ribbon modes, using the tabulated values of $\zeta_n$ and $A_n$ from Ref.~\citenum{christensen2017from}.
Graphene's conductivity is a sum of intra- and interband contributions, \ie $\sigma\left(\omega\right)=\sigma_{\text{intra}}\left(\omega\right)+\sigma_{\text{inter}}\left(\omega\right)$, contributing spectral features of Drude- and Landau-damping kinds, respectively~\cite{Note100}
\footnotetext[100]{%
    The conductivity expression of \cref{eq:conductivity_inter} neglects the temperature-dependence of the interband term because the usual expressions for the temperature-dependence of the interband term~\cite{koppens2011graphene, falkovsky2007space} are incompatible with the auto-differentation tools used here.
    This omission, however, has negligible impact on the overall response of the network due to the weakness of the temperature-dependence.
}
\begin{subequations}\label[subequations]{eqs:conductivity}
\begin{align}
    \label{eq:conductivity_intra}
     \sigma_{\text{intra}}(\omega)
     &=
      \frac{2 \iu e^2 \kb T}{\pi \hbar^2 (\omega + \iu \gamma)}\ln\left[2\cosh\left(\frac{\Ef}{2\kb T}\right)\right],
     \\
     \label{eq:conductivity_inter}
     \sigma_{\text{inter}}(\omega) 
     &=
      \frac{\iu e^2}{4 \pi \hbar} \ln \left[ \frac{2 |\Ef| - \hbar (\omega + \iu \gamma)}{2 |\Ef| + \hbar (\omega + \iu \gamma) } \right],
\end{align}
\end{subequations}
where $\kb$ denotes the Boltzmann constant, $e$ the electronic charge, $\hbar$ the reduced Planck's constant, $\gamma$ a relaxation rate, and $T$ the ambient temperature.

\begin{figure}
    \includegraphics[scale=1]{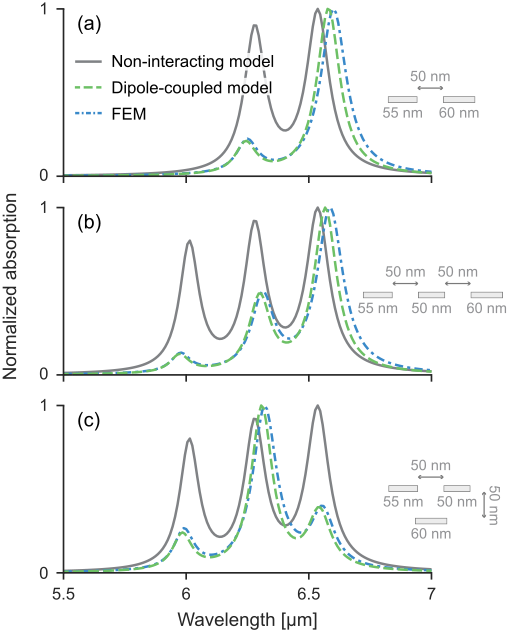}
    \caption{%
        \textbf{Validation of the network model.}
        (a)~two dissimilar elements in a planar configuration.
        (b)~three dissimilar elements in a planar configuration.
        (c)~three dissimilar elements in a stratified two-layer configuration. 
        The dipole-coupled model (dashed green) matches well with finite-element simulations (FEM) in Comsol (dashed-dotted blue). 
        In contrast, a simple, non-interacting analysis that ignores inter-ribbon coupling (solid gray) models the response inadequately.
    }
    \label{fig:2}
\end{figure}
\begin{figure}
    \includegraphics[scale=1]{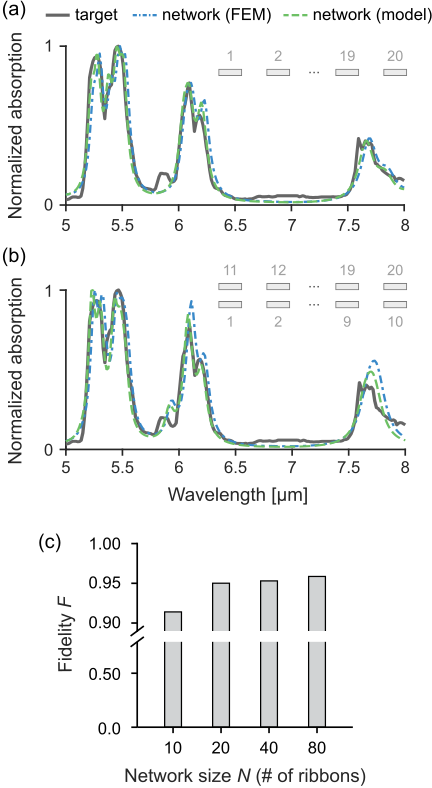}
    \caption{%
        \textbf{Spectrally mimicking a single complex target.}
        The network is optimized to mimic the absorption profile of nitric oxide gas (NO) in a planar (a) and a stratified configuration (b).
        The model (dashed green), the target spectrum (solid purple), and the finite-element validation (dash-dotted blue) show good agreement.
        (c)~Fidelity of matching (\cref{eq:F-single}) as a function of the network size.
    }
    \label{fig:3}
\end{figure}

To model a network of nanoresonators, we consider coupling of multiple elements via their induced dipole moments, following the coupled dipole approach.
In particular, we consider a collection of $i = 1, \ldots, N$ ribbons, each of size $W_i$, Fermi energy $E_{\text{F}i}$, and associated polarizability $\alpha_i$.
For definiteness, we assume each ribbon to be of finite width along $x$, infinitely extended along $y$, infinitesimal along $z$, and define $\rv_\perp \equiv (x,z)$ to denote the non-invariant coordinates.
The induced dipole moment $p_i$ of the $i$-th ribbon is proportional to the electric field \emph{not} originating from the $i$-th ribbon itself, such that:
\begin{equation}
    \label{eq:induced_dipole}
    p_i 
    = \varepsilon_0\varepsilon_{\text{B}}\alpha_i(\omega) E_i',
\end{equation}
where $E_i'$ denotes the ribbon-averaged sum of the in-plane components of any external field, $E_0$, and the in-plane components of the induced fields of all other ribbons $E_{j\neq i}$:
\begin{equation}
    \label{eq:driving_fields}
    E_i' 
    \equiv
    \frac{1}{W_i}\int_{i}  {E}_0(\rv_\perp) \dde{x}
    +
    \frac{1}{W_i}\int_{i}\sum_{j \neq i} {E}_j(\rv_\perp) \dde{x},
\end{equation}
with $\int_i \dde{x}$ denoting integration over the $i$-th ribbon's in-plane extent.
In turn, the dipole $p_i$ induces a field $E_i(\rv_\perp)$ originating from the $i$-th ribbon, whose in-plane components are
\begin{equation}
    \label{eq:induced_field}
    E_i(\rv_\perp)
    =
    \omega^2\mu_0 G(\rv_\perp,\rv_{\perp i})p_i,
\end{equation}
with $G(\rv_\perp,\rv_{\perp i})$ denoting the $y$-integrated $xx$-component of the nonretarded free-space Green function and $\rv_{\perp i}$ the center coordinate of the $i$-th ribbon.
Combining \cref{eq:induced_dipole,eq:driving_fields,eq:induced_field}, we obtain a system of coupled equations, which we cast as:
\begin{equation}
    \label{eq:coupled_equations}
    (\boldsymbol{1} - k_{\text{B}}^2 \boldsymbol{\alpha} \mathbf{G}) \mathbf{p}
    =
    \varepsilon_0 \varepsilon_{\text{B}} \boldsymbol{\alpha} \Ev_0,
\end{equation}
with wave number $k_{\text{B}}^2 \equiv \varepsilon_{\text{B}}\omega^2/c^2$, dipole moments $\mathbf{p} \equiv [p_1, \ldots, p_N]$, ribbon-averaged external fields $\mathbf{E}_0 \equiv \big[W_1^{-1}\int_1 E_0(\rv_{\perp})\dde{x},\ldots, W_N^{-1} \int_N E_0(\rv_{\perp})\dde{x}\big]$, diagonal ``bare'' polarizability matrix $\boldsymbol{\alpha} \equiv  \mathop{\mathrm{diag}}(\alpha_1, \ldots, \alpha_N)$, and Green function matrix $(\mathbf{G})_{ij} \equiv (1 - \delta_{ij})\bar{G}_{ij}$ whose elements give the dipole-coupling between the $i$-th and $j$-th ribbon (\cref{app:green_elements}):
\begin{align}
    \bar{G}_{ij}
    &\equiv
    \frac{1}{W_i}\int_i G(\rv_\perp,\rv_{\perp j})\dde{x}
    \nonumber\\
    &=
    \frac{2}{\pi k_{\text{B}}^2}\frac{4(x_{ij}^2-z_{ij}^2)-W_i^2}{\big[(W_i+2x_{ij})^2+4z_{ij}^2)\big]\big[(W_i-2x_{ij})^2+4z_{ij}^2\big]}.
    \label{eq:green_elements}
\end{align}
where $x_{ij} \equiv x_i - x_j$ and $z_{ij} \equiv z_i - z_j$ are the center-to-center $x$- and $z$-separations, respectively, between the $i$-th and $j$-th ribbon.

It is convenient to express the solution to  \cref{eq:coupled_equations} in terms of a ``dressed'' or effective polarizability $\boldsymbol{\alpha}_{\text{eff}} \equiv (\boldsymbol{1}-k_\text{B}^2\boldsymbol{\alpha}\mathbf{G})^{-1}\boldsymbol{\alpha}$ such that the induced dipole amplitudes and ribbon-averaged total fields are $\mathbf{p} = \varepsilon_{0}\varepsilon_{\text{B}}\boldsymbol{\alpha}_{\text{eff}}\Ev_0$ and $\Ev' = (\boldsymbol{1}+k_{\text{B}}^2\mathbf{G}\boldsymbol{\alpha}_{\text{eff}})\Ev_0$.
Similarly, we can express the main quantity of interest in our present case---the total absorption $P_{\text{abs}}$ due to a monochromatic and uniform external field---in terms of $\boldsymbol{\alpha}_{\text{eff}}$ and $\mathbf{G}$, \cf:
\begin{align}
    P_{\text{abs}}
    &=
    \tfrac{1}{2}\Re\int
    \mathbf{J}^*(\rv_\perp) \cdot\mathbf{E}(\rv_\perp) \dde{V}
    =
    -\tfrac{1}{2}\omega\Im 
    \big(\mathbf{p}^\dagger\Ev'\big)
    \nonumber\\
    &= 
    -\tfrac{1}{2}\varepsilon_{0}\varepsilon_{\text{B}}\omega\Im\big[
    \big(\boldsymbol{\alpha}_{\text{eff}}\Ev_0)^\dagger
    (\boldsymbol{1}+k_{\text{B}}^2\mathbf{G}\boldsymbol{\alpha}_{\text{eff}})\Ev_0\big],
\end{align}
where we have used that the induced current $\mathbf{J}(\rv_\perp)$ is a sum over dipole terms $\iu\omega p_i\mathbf{\hat{x}}$.

\Cref{fig:2} shows the implementation of the semi-analytical dipole-coupled model.
We emphasize two key points.
First, the dipole-coupled model closely matches the finite-element simulations in COMSOL Multiphysics, which capture the complete interaction but are much slower and computationally demanding due to their discretizing nature.
Second, the dipole-coupled model can account for coupling that is not captured in the much simpler, analytic, treatment of non-coupled ribbons.

We analyze the response arising from the interaction between a small number of ribbons (two and three) in both planar and stratified configurations.
To maintain generality, we assume that the ribbons are in a vacuum background and not on a particular substrate (the effects of a substrate can be accounted for through a modified Green function). The Fermi energies of ribbons are all set to 0.35~eV, with $\gamma = 3\,\text{meV}$ and $T = 300\,\text{K}$.
\Cref{fig:2}a shows the normalized absorption for the case of two elements of dissimilar size.
We observe a correct prediction of a stronger resonance at 6.6~µm and a suppressed resonance at 6.3~µm.
This contrasts with the inaccurate prediction from the non-interacting analysis (solid grey) which is inadequate for modeling the system.
For the case of three dissimilar elements arranged in a planar configuration, \cref{fig:2}b demonstrates a very good agreement between the dipole-coupled semi-analytical model (solid green) and the finite element simulations (dashed blue).
The absorption response comprises the resonances from the three ribbons, but the interaction between the ribbons modulates the overall magnitudes, resulting in a noticeably weaker absorption of two narrower ribbons.
Finally, the semi-analytical model also works for non-planar arrangements.
\Cref{fig:2}c depicts an example of a stratified three-ribbon configuration.
For this configuration, the Green function is modified to account for the vertical coupling between the ribbons.
Unlike the two previous planar cases, we observe the strongest absorption at the wavelength corresponding to the resonance of the ribbon with the intermediate width (55~nm), with the other two absorption peaks suppressed.

In all cases in \cref{fig:2}, we see a very good match between the semi-analytical model and the finite-element simulations.
The frequencies, linewidths, and relative strengths of the absorption peaks are predicted with good accuracy, with the semi-analytical model showing a slight blue-shift (e.g., $ {\mathrm{\Delta\lambda}}/{\lambda} \sim 0.004$ for the 6.6~µm peak in \cref{fig:2}a).
In contrast, the non-interacting analysis that ignores the coupling between the elements models the system response poorly.

\begin{figure*}
     \includegraphics[width=.925\textwidth]{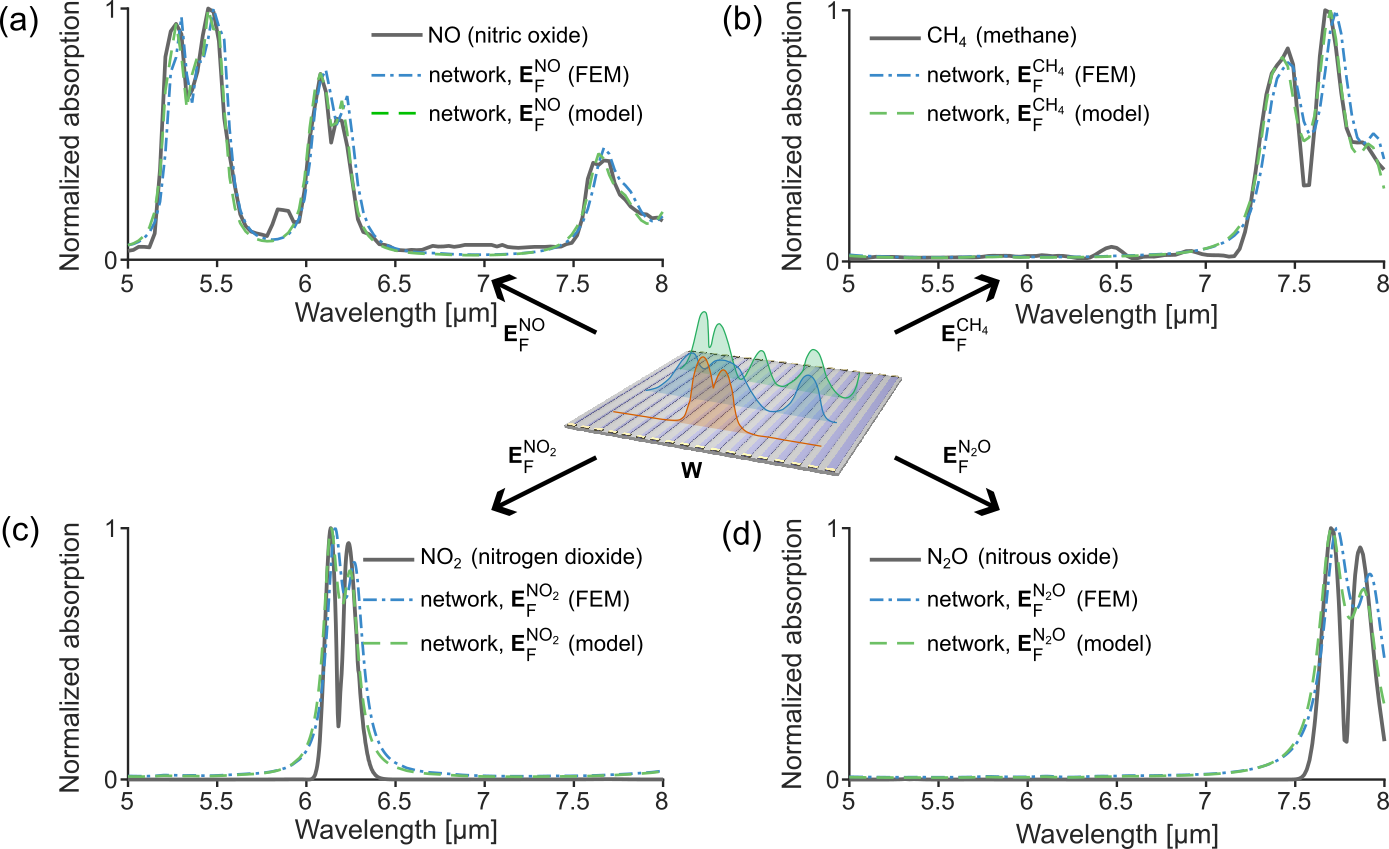}
    \caption{%
        \textbf{A single network reconfigures between four complex absorption profiles.}
        To mimic the $M=4$ target spectral profiles, we design a single optimal geometric arrangement and $M$ optimal voltage inputs.
        A 4-target mean fidelity of $F=0.91$ is achieved.
        The lines correspond to the model prediction (dashed green), target spectra (solid purple), and the finite-element simulation (dash-dotted blue).
    }
    \label{fig:4}
\end{figure*}

\section{Results}
Having established the coupled elements as building-blocks for the network, we proceed to show how the network can be designed to mimic a complex spectrum.
As an illustration of a complex spectral profile, we select a target that resembles the absorption of nitric oxide (NO).
\Cref{fig:3} shows the low resolution NO absorption directly obtained from the NIST Chemistry WebBook~\cite{Linstrom2001-vm}.
The NO absorption profile contains multiple broad and narrow spectral features in the 5\,--\,8~µm range, rendering it a suitable target for exploring the potential of our approach.

The design of the mimicking network involves determining the optimal network geometry (i.e., ribbon widths and arrangements) and the optimal input signals (i.e., gate-tunable ribbon Fermi levels).
Mathematically, we seek to maximize spectral matching over a bandwidth $\mathrm{\Delta\lambda}$, which we quantify by a fidelity metric 
\begin{equation}
    F 
    =
    1 - \sqrt{\frac{1}{\Delta \lambda} \int_{\lambda_{\text{min}}}^{\lambda_{\text{max}}}
    \!\!
    \big[A_{\text{target}}(\lambda)-A_{\text{network}}(\lambda;\mathbf{W},\mathbf{E}_{\text{F}})\big]^2\dde{\lambda}}.
    \label{eq:F-single}
\end{equation}
This metric is a function of input vectors or ribbon widths $\mathbf{W}=[W_1,\ \ldots,\ W_N]$, and energies $\mathbf{E}_{\text{F}} =[E_{\text{F}1},\ \ldots,\ E_{\text{F}N}]$ within the network.
Here, $\mathrm{\Delta\lambda}$ denotes the bandwidth of the target absorption spectrum $A_{\text{target}}(\lambda)$ and $A_{\text{network}}\left(\lambda\right)$ denotes the spectrum of the network.
Both spectra are scaled and normalized to unity, similar to outputs observed in spectroscopic measurements.
The fidelity function, $F$, assesses the average spectral separation (or spectral overlap) between the target spectrum and the network spectrum.
The limit of $F\rightarrow1$ indicates a perfect overlap between the two normalized spectra.

Efficient maximization of fidelity $F$ over the $N+N$ parameter space of ribbon widths and energies is made possible by the fact that gradients $\partial F/\partial W_i$ and $\partial F/\partial E_{\text{F}i}$ can be computed very efficiently via auto-differentiation \cite{Zygote.jl-2018}.
By combining gradient-based optimization (facilitated by auto-differentiation) with the speed of the semi-analytical model, we can explore a large design space of many ribbons and initial conditions.
To mimic the NO spectrum, we start by generating a set of initial conditions for the vector $(\mathbf{W},\mathbf{E}_{\text{F}})$ using low-discrepancy Sobol sequences which help to achieve a well-distributed set of points within the $2N$-dimensional hypercube~\cite{Joe2003-pm}.
For each initial condition, the network is optimized using the method of moving asymptotes (MMA) algorithm~\cite{Svanberg2002-ay} (accessed through the NLopt package~\cite{Johnson2014-rd}).
For all $E_\text{Fi}$ in the $\mathbf{E}_\text{F}$ set, we impose Fermi energy bounds $0.1\ \mathrm{eV}\le E_{\text{F}i} \le 0.5\ \mathrm{eV}$ and a minimum ribbon width of $40\ \mathrm{nm} \le W_i$.
\Cref{fig:3}a shows the optimized result for the network consisting of $N=20$ ribbons (associated ribbon parameters are listed in \cref{app:network-params}).
We observe a very close match between the target spectrum (solid purple) and the network spectrum calculated using the dipole-coupled model (dashed green) and validated using a finite-element simulation (dotted blue).
Importantly, the network does not need to be in a single planar layer.
Stratified, multi-layer, configurations can be beneficial to enhance the per-area response and minimize device size.
\Cref{fig:3}b demonstrates that the same optimization can be applied to a two-level configuration, yielding similarly very good performance.

\Cref{fig:3}c compares the achieved fidelity $F$ for networks of varying sizes.
For smaller networks ($N=10$ and $20$), a Sobol set of $1024$ initial conditions is generated; for larger networks ($N=40$ and $80$), the results from smaller networks ($N=20$ and $40$, respectively) are used to inform pre-optimization initial conditions.
Interestingly, even a smaller network of $N=10$ elements can perform decently well ($F=0.91$).
Its spectrum is shown in \cref{app:10-element-result}.
The largest improvement is seen in the transition from an $N=10$ to an $N=20$ network ($F=0.95$).
Further enlarging the network continues to improve fidelity, but with diminishing benefits.
This indicates that the intrinsic linewidth constraints become a limiting factor for spectral matching.
We elaborate on this in \cref{sec:discussion}.

Finally, we explore how a single network of fixed geometry can mimic multiple different complex spectral profiles.
We select spectral absorption profiles of nitric oxide (NO), methane (CH$_4$), nitrogen dioxide (NO$_2$), nitrous oxide (N$_2$O) as targets (\cref{fig:4}).
For NO, the target spectrum is directly extracted from the database; for CH$_4$, N$_2$O, and NO$_2$, a spectral envelope is formed to mimic spectral trends, where the envelopes are created using spline interpolation over local maxima separated by at least 100 nm~\cite{Linstrom2001-vm,Gordon2017-av}.
Generally, for $M$ targets, the design problem is to find one optimal network geometry ($\mathbf{W}=[W_1,\ldots,W_N]$) and $M$ optimal Fermi energy sets ($\mathbf{E}_{\text{F}}^{(k)}=[E_\text{F1}^{(k)},\ldots,E_\text{FN}^{(k)}]$, for $k=1,\ldots, M$) that, together, maximize the overall fidelity of spectral mimicking.
Mathematically, we express the $M$-target fidelity function $F_{\!M}$ as the average of the individual target fidelities: 

\begin{align}
    F_{\!M} &= \big\langle F^{(k)} \big\rangle_{k\,=\,1}^M,
    \\
    F^{(k)}
    &=
    1 -
    \sqrt{\frac{1}{\Delta \lambda^{(k)}}\!
    \int_{\lambda_{\text{min}}^{(k)}}^{\lambda_{\text{max}}^{(k)}}
    \!\!
    \big[A_{\text{target}}^{(k)}(\lambda) - A_{\text{network}}(\lambda;\mathbf{W},\mathbf{E}_{\text{F}}^{(k)})\big]^2
    \dde{\lambda}},
    \nonumber
\end{align}
where $\langle\cdot\rangle_{k\,=\,1}^M$ denoting the average over $M$ distinct targets. For a network of $N$ elements (e.g., ribbons), the dimension of the optimization space is $\left(M+1\right)N$ to account for $M\times N$ input signals (\ie Fermi energies $E_{\text{F}i}^{(k)}$), and $N$ geometric parameters (\ie ribbon widths $W_i$).
As before, the auto-differentiable implementation makes it possible to calculate all gradients with respect to both input signals $\partial F_{\!M}/\partial E_{\text{F}i}^{(k)}$ and the geometry $\partial F_{\!M}/\partial W_i$ very efficiently.

\Cref{fig:4} shows the four spectra of the network designed to replicate the four example gases.
The network consists of $N=20$ ribbons, in a planar arrangement, with the initial optimization parameters informed by results from \cref{fig:3}.
We observe a relatively high net fidelity value of $F_{\!M}=0.91$.
The spectra of NO and CH$_4$ show a better match ($F^{\text{NO}}=0.95$ and $F^{\text{CH}_4}=0.96$) relative to the spectra of NO$_2$ and N$_2$O ($F^{\text{NO}_2}=0.87$ and $F^{\text{N}_2\text{O}}=0.86$).
This is attributed to the presence of narrower features in NO$_2$/N$_2$O spectra whose linewidths become too small for the network of this size and intrinsic graphene properties to accurately resolve.
The optimized parameters of the network, the geometry and the four sets of input signals, are given in \cref{app:network-params}.
Overall, this example network showcases the promising potential of the coupled system to synthesize (and reconfigure between) complex spectral profiles.

\section{Discussion and Conclusion}
\label{sec:discussion}
We envision several relevant extensions of this work.
In our analysis, we focused on graphene elements in a vacuum background to maintain generality.
In a practical implementation, the effects of a substrate can be taken into account through a modified Green function.
The introduction of a substrate, especially a nano-patterned or layered substrate, provides an additional opportunity to create more complex resonances (e.g., Fano resonances) and improve matching fidelity.
Further, we note that the intrinsic material properties of graphene (e.g., damping $\gamma$) constrains the minimal linewidth of the spectral response and that higher carrier mobility would allow mimicking narrower resonances. 
The model can be generalized beyond dipole-coupling which might be relevant for ribbons with very small gaps between them.
We have here analyzed networks in both planar and stratified configurations; in a practical device, one would want to consider the trade-off between the network density and the complexity of the gating architecture.

In summary, we have theoretically demonstrated how a metamaterial network comprising coupled, gate-tunable, 2D-material resonators in graphene can mimic and reconfigure between multiple absorption spectra. 
By deriving a semi-analytical dipole-coupled network model and implementing it through auto-differentiation, we enabled a scalable design of high-dimensional networks with many elements and input signals.
Our investigation shows that such networks can be designed to accurately match multiple complex spectral profiles, such as those resembling gases. 
Among known 2D materials, graphene has the strongest material-specific optical response in the longwave-IR/thermal-IR, independent of shape or structuring~\cite{Miller2017-gb}.
However, our proposed approach is general and applicable to other (gate-)tunable 2D materials or coupled resonators in structures where tunability is achieved by other means, such as deformation or phase change.
The underlying concept could facilitate the design of optical materials and platforms with excellent reconfigurability with applications in multi-spectral systems.

\appendix
\renewcommand{\figurename}{Supplementary Figure}
\renewcommand{\tablename}{Supplementary Table}

\section{Elements of Green function matrix}
\label{app:green_elements}
We derive \cref{eq:green_elements} starting from the definition of the $y$-integrated $xx$-component of the free-space Green function (embedded in a medium of permittivity $\varepsilon_{\text{B}}$).
The starting point is the nonretarded, uniform-medium Green tensor $\tensor{\mathbf{G}}(\rv,\rv') = \tensor{\mathbf{G}}(\tilde{\rv}) = \big[\hat{\tilde{\rv}}\otimes\hat{\tilde{\rv}}-\tensor{\boldsymbol{1}}\big]/({4\pi k_{\text{B}}^2|\tilde{\rv}|^3})$ where $\tilde{\rv} \equiv \rv - \rv'$.
Taking the $xx$-component and integrating out the invariant $y$-direction, we obtain the induced field at $\rv_{\perp} = (x,z)$ from a $y$-extended dipole ``line'' at fixed $\rv_\perp' = (x',z')$:
\begin{align}
    G(\rv_\perp,\rv'_\perp)
    &=
    G(\tilde{\rv}_\perp)
    \nonumber\\
    &\equiv
    \int_{-\infty}^{\infty} G_{xx}(\tilde{\rv})\dde{\tilde{y}}
    \nonumber\\
    &=
    \frac{1}{4\pi k_{\text{B}}^2}
    \int_{-\infty}^{\infty}
    \frac{3\tilde{x}^2}{|\tilde{\rv}|^5}
    - \frac{1}{|\tilde{\rv}|^3}\dde{y}
    \nonumber\\
    &=
    \frac{1}{2\pi k_{\text{B}}^2}
    \Bigg(
    \frac{2\tilde{x}^2}{|\tilde{\rv}_\perp|^4}
    - \frac{1}{\tilde{\rv}_\perp^2}
    \Bigg).
    \label{eq:green_line}
\end{align}
Next, to obtain \cref{eq:green_elements}, we average \cref{eq:green_line} over the extent of a ribbon at $i$ while setting $\rv_\perp' = \rv_{\perp j}$:
\begin{align}
    \bar{G}_{ij}
    &= 
    \frac{1}{W_i}\int_i 
    G(\rv_\perp,\rv'_\perp)
    \dde{x}
    \nonumber\\
    &=
    \frac{1}{2\pi k_{\text{B}}^2}
    \frac{1}{W_i}
    \int_i 
    \Bigg(
    \frac{2\tilde{x}^2}{|\tilde{\rv}_\perp|^4}
    - \frac{1}{\tilde{\rv}_\perp^2}
    \Bigg)\dde{x}
    \nonumber\\
    &=
    \frac{1}{2\pi k_{\text{B}}^2}
    \frac{1}{W_i}
    \int_{-W_i/2}^{W_i/2}
    \Bigg(\frac{2(\tilde{x}+x_{ij})^2}{\Big[\!\!\sqrt{(\tilde{x}+x_{ij})^2 + z_{ij}^2}\Big]^4}
    \nonumber\\&\hspace{4cm}
    - \frac{1}{\Big[\!\!\sqrt{(\tilde{x}+x_{ij})^2 + z_{ij}^2}\Big]^2}
    \Bigg)\dde{\tilde{x}}
    \nonumber\\
    &= 
    \frac{2}{\pi k_{\text{B}}^2}\frac{4(x_{ij}^2-z_{ij}^2)-W_i^2}{\big[(W_i+2x_{ij})^2+4z_{ij}^2)\big]\big[(W_i-2x_{ij})^2+4z_{ij}^2\big]}.
\end{align}

\section{Network parameters}
\label{app:network-params}
We tabulate the network parameters used for \cref{fig:3,fig:4} of the main text in 
\href{tab:fig3a_params}{Supplementary Tables~\ref{tab:fig3a_params}}, \ref{tab:fig3b_params}, and \ref{tab:fig4_params}.

\begin{table}[!htbp]
    \begin{tabular}{rcc}
        \toprule
        $i$ & $W_i$ [nm] & $E_i$ [eV]\\
        \midrule
        1  & 44.2 & 0.151\\
        2  & 40.2 & 0.187\\
        3  & 52.7 & 0.357\\
        4  & 56.3 & 0.100\\
        5  & 57.4 & 0.100\\
        6  & 50.0 & 0.420\\
        7  & 50.4 & 0.121\\
        8  & 54.8 & 0.482\\
        9  & 53.5 & 0.211\\
        10 & 50.4 & 0.333\\
        11 & 55.7 & 0.499\\
        12 & 59.0 & 0.125\\
        13 & 60.0 & 0.105\\
        14 & 47.3 & 0.331\\
        15 & 50.6 & 0.140\\
        16 & 52.2 & 0.427\\
        17 & 57.4 & 0.124\\
        18 & 41.3 & 0.442\\
        19 & 48.6 & 0.420\\
        20 & 56.7 & 0.250\\
        \bottomrule
    \end{tabular}
    \caption{Network parameters for \cref{fig:3}a.
    Center-to-center distances between ribbons are set to 120~nm.}
    \label{tab:fig3a_params}
\end{table}

\begin{table}[!htbp]
    \begin{tabular}{rcc}
        \toprule
        $i$ & $W_i$ [nm] & $E_i$ [eV]\\
        \midrule
        1  & 50.4 & 0.138\\
        2  & 41.3 & 0.101\\
        3  & 54.1 & 0.166\\
        4  & 51.1 & 0.449\\
        5  & 44.4 & 0.105\\
        6  & 51.9 & 0.427\\
        7  & 50.2 & 0.350\\
        8  & 53.3 & 0.377\\
        9  & 47.3 & 0.219\\
        10 & 56.2 & 0.467\\
        11 & 50.4 & 0.333\\
        12 & 45.8 & 0.211\\
        13 & 46.4 & 0.390\\
        14 & 43.1 & 0.170\\
        15 & 44.8 & 0.109\\
        16 & 41.9 & 0.392\\
        17 & 50.3 & 0.462\\
        18 & 54.8 & 0.201\\
        19 & 41.2 & 0.100\\
        20 & 41.6 & 0.196\\
        \bottomrule
    \end{tabular}
    \caption{Network parameters for \cref{fig:3}b.
    Center-to-center distances between ribbons are set to 120~nm, and separation between two layers is set to 80~nm.}
    \label{tab:fig3b_params}
\end{table}

\begin{table}[!htbp]
    \centering
    \begin{tabular}{rccccc}
        \toprule
        $i$ & $W_i$ [nm] & $E_i^{\text{NO}}$ [eV] & $E_i^{\text{CH}_4}$ [eV] & $E_i^{\text{N}_2\text{O}}$ [eV] & $E_i^{\text{NO}_2}$ [eV]\\
        \midrule
        1 & 44.1 & 0.151 & 0.491 & 0.145 & 0.110\\
        2 & 40.3 & 0.189 & 0.120 & 0.143 & 0.160\\
        3 & 52.7 & 0.357 & 0.235 & 0.147 & 0.352\\
        4 & 56.2 & 0.105 & 0.050 & 0.144 & 0.168\\
        5 & 57.5 & 0.103 & 0.176 & 0.254 & 0.209\\
        6 & 50.0 & 0.420 & 0.142 & 0.230 & 0.190\\
        7 & 50.4 & 0.118 & 0.097 & 0.135 & 0.195\\
        8 & 54.8 & 0.482 & 0.260 & 0.233 & 0.352\\
        9 & 53.4 & 0.217 & 0.064 & 0.236 & 0.127\\
        10 & 49.8 & 0.330 & 0.050 & 0.123 & 0.108\\
        11 & 55.8 & 0.500 & 0.247 & 0.245 & 0.370\\
        12 & 58.9 & 0.119 & 0.277 & 0.126 & 0.187\\
        13 & 60.0 & 0.105 & 0.186 & 0.170 & 0.123\\
        14 & 46.9 & 0.330 & 0.215 & 0.147 & 0.175\\
        15 & 50.6 & 0.141 & 0.245 & 0.173 & 0.108\\
        16 & 52.2 & 0.427 & 0.149 & 0.148 & 0.350\\
        17 & 57.5 & 0.125 & 0.237 & 0.101 & 0.145\\
        18 & 41.0 & 0.456 & 0.121 & 0.099 & 0.171\\
        19 & 48.5 & 0.420 & 0.238 & 0.140 & 0.318\\
        20 & 56.9 & 0.252 & 0.268 & 0.248 & 0.119\\
        \bottomrule
    \end{tabular}
    \caption{Network parameters for \cref{fig:4}. Center-to-center distances between ribbons are set to 120~nm.}
    \label{tab:fig4_params}
\end{table}

\newpage
\section{Small network (10-element) result}
\label{app:10-element-result}

\begin{figure}[!htbp]
    \includegraphics[width= 0.925\columnwidth]{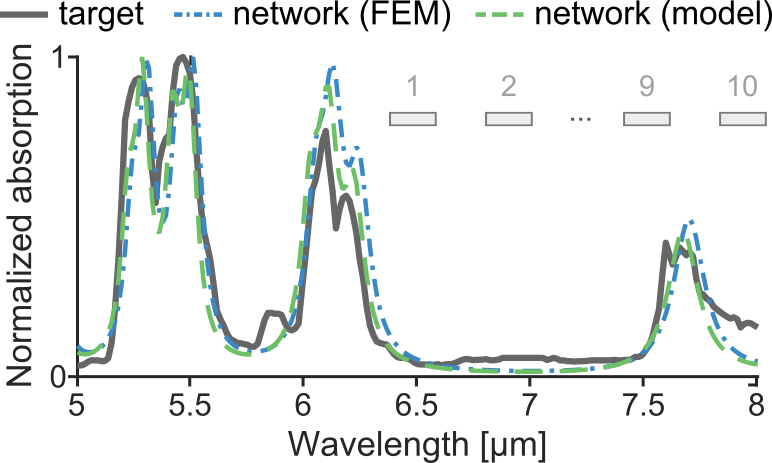}
    \caption{%
        \textbf{Spectrally mimicking a single complex target with a small network}
        The 10-ribbon network is optimized to mimic the absorption profile of NO.
    }
    \label{fig:10-ribbon_res}
\end{figure}

We present the result of a 10-ribbon network optimization in \href{fig:10-ribbon_res}{Supplementary Fig.~\ref{fig:10-ribbon_res}}, which shows the network's absorption profile for NO. The associated network parameters $(\mathbf{W},\mathbf{E}_{\text{F}})$ are listed in \href{tab:10-ribbon_params}{Supplementary Table~\ref{tab:10-ribbon_params}}.

\begin{table}[!htbp]
    \centering
    \begin{tabular}{rcc}
        \toprule
        $i$ & $W_i$ [nm] & $E_i$ [eV] \\
        \midrule
        1 & 42.6 & 0.289 \\
        2 & 42.4 & 0.445 \\
        3 & 47.2 & 0.394 \\
        4 & 46.3 & 0.328 \\
        5 & 56.6 & 0.470 \\
        6 & 40.8 & 0.290 \\
        7 & 57.4 & 0.203 \\
        8 & 48.2 & 0.430 \\
        9 & 54.1 & 0.240 \\
        10 & 46.5 & 0.427 \\
        \bottomrule
        \end{tabular}
    \caption{Network parameters for \href{fig:10-ribbon_res}{Supplementary Fig.~\ref{fig:10-ribbon_res}}. Center-to-center distances between ribbons are set to 120 nm.}
    \label{tab:10-ribbon_params}
\end{table}

\FloatBarrier
\bibliographystyle{apsrev4-2-longbib}

\end{document}